\documentstyle[aps,eqsecnum,epsf]{revtex}
\begin{document}
\thispagestyle{empty}
%\hspace{-15mm}
%\leftline{\epsfbox{mark.eps}}  
%\vspace{-13.0mm} % for revtex
%\vspace{-9.3mm} % for article, article  11pt
%\vspace{-10.3mm} % for article 12pt, revtex preprint style
%\hspace{18mm}

{\baselineskip-4pt
\font\yitp=cmmib10 scaled\magstep2
\font\elevenmib=cmmib10 scaled\magstep1  \skewchar\elevenmib='177
\leftline{\baselineskip20pt
%\hspace{12mm} % for revtex
%\hspace{15mm} % for article, revtex preprint style
\vbox to0pt
   { {\yitp\hbox{Osaka \hspace{1.5mm} University} }
%\vspace{-4mm} % for revtex preprint style
     {\large\sl\hbox{{Theoretical Astrophysics}} }\vss}}

%{\baselineskip-4pt
%\font\yitp=cmmib10 scaled\magstep2
%\font\elevenmib=cmmib10 scaled\magstep1  \skewchar\elevenmib='177
%\leftline{\baselineskip20pt\vbox to0pt
%   { {\yitp\hbox{Osaka \hspace{1.5mm} University} }
%     {\large\sl\hbox{{Theoretical Astrophysics}} }\vss}}

\rightline{\large\baselineskip20pt\rm\vbox to20pt{
\baselineskip14pt
\hbox{OU-TAP 57}
\hbox{astroph/9702xxx}
\vspace{2mm}
\hbox{February, 1997}\vss}}%
%}
\vskip3cm
\begin{center}{\large\bf
Wall fluctuation modes and tensor CMB anisotropy in
open inflation models}
\end{center}
%\vfill
\vspace{0.5cm}
\begin{center}
{\large Misao Sasaki,\footnote{Electronic address:
 misao@vega.ess.sci.osaka-u.ac.jp}
Takahiro Tanaka\footnote{Electronic address:
 tama@vega.ess.sci.osaka-u.ac.jp}
and 
Yoshihiro Yakushige\footnote{Electronic address:
 yakusige@vega.ess.sci.osaka-u.ac.jp}}\\
\vspace{0.5cm}
{Department of Earth and Space Science, \\
Osaka University,~Toyonaka 560,~Japan}\\
\vspace{0.4cm}
\end{center}
%\vfill
\begin{abstract}
We calculate the spectrum of large angle cosmic microwave background (CMB)
 anisotropies due to quantum fluctuations of the gravitational wave
modes in one-bubble open inflation models.
We find the bubble-wall fluctuation modes, which had been thought to
exist discretely in previous analyses, are actually contained in the
continuous spectrum of gravitational wave modes when the gravitational
coupling is correctly taken into account. Then we find that the spectrum
of the tensor CMB anisotropy can be decomposed into the part due to
the wall fluctuation modes and that due to the
usual gravitational wave modes in a way which is almost
model-independent, even when the gravitational coupling is strong.
We also discuss observational constraints on the model parameters.
We find that an appreciable portion of the parameter space is excluded
but the remaining allowable region is still wide enough
to leave the one-bubble scenario viable.
\end{abstract}

%%%%%%%%%%%%%%%%%%%%%%%%%%%%%%%%%%%%%%%%%%%%%%%%%%%%%%%%%%%%%%%
\section{Introduction}
The inflationary universe scenario is one of the most successful 
scenarios that can explain the homogeneity and isotropy of the
universe on very large scales as well as the inhomogeneous structure
on smaller scales such as galaxies and clusters of galaxies.
The standard model of inflation predicts that our universe is spatially 
flat with $\Omega_0=1$.
However, there are increasing observational evidences that indicate 
$\Omega_0<1$\cite{Omega}.
Consequently the standard model of inflation
needs to be modified to account for $\Omega_0<1$. As one of such
 modifications, the one-bubble open inflationary scenario has attracted
much attention recently. 

The basic idea of the one-bubble open inflationary 
universe scenario was proposed by Gott III\cite{Gott} 
fifteen years ago and was revived recently
by several authors\cite{YST95,BGT,Linde}.
In this scenario, there are two stages of inflation.
The universe is in the false vacuum initially and this first stage of
inflation is assumed to last long enough so that the universe becomes
sufficiently homogeneous and isotropic and becomes well-approximated by
a pure de Sitter space. Then the nucleation of a vacuum bubble occurs
through quantum tunneling. This process is described by a Euclidean
bounce solution, which is a non-trivial
classical solution of the field equation in Euclidean 
spacetime having $O(4)$-symmetry\cite{Col,ColDeL}.
Then the expanding bubble after nucleation is described 
by the classical solution obtained by analytic 
continuation of the bounce solution to Lorentzian spacetime.
Owing to the $O(4)$-symmetry of the bounce solution, the
expanding bubble has $O(3,1)$-symmetry. 
This implies that the system is homogeneous and isotropic
on the hyperbolic time slicing inside the bubble and that 
the nucleation of a bubble can be regarded as the creation of 
an open universe\cite{Gott,MTYY93}.
Then it is assumed that the vacuum energy inside the bubble is
non-zero and the second stage of inflation commences.
One assumes this inflation is of slow rollover type and it lasts
just enough to make the present day density parameter $\Omega_0$
appreciably smaller than unity.

There are essentially two types of one-bubble open inflation models;
single field models\cite{YST95,BGT} and
two field models\cite{Linde,AmeBac,GreLid}.
 In the former case
the false vacuum decay and the subsequent stage of inflation
inside the bubble are both governed by a single scalar field, while in
the latter case they are driven by two different fields.

In both cases, the quantum fluctuations of the inflaton field at the
second stage of inflation give rise to curvature perturbations of the
universe which account for the present large scale structure of the
universe. The spectrum of these fluctuations was calculated and
the resulting CMB anisotropy was evaluated by several
authors\cite{YST95,BGT,YamBun,Garriga,Bellido,YST96,Cohn}.
Through these works, it was clarified that there could appear discrete
modes which have a coherence scale greater than the spatial curvature
scale of the universe (the so-called de Sitter supercurvature modes)
\cite{STY95,YST96}.
Their existence could be harmful because of their large contribution
to the CMB anisotropy on largest angular scales\cite{YamBun}. 
In fact, the simplest two field model which assumes two decoupled
scalar fields\cite{Linde} was shown to be unsuccessful in this
respect\cite{ST96}.
Fortunately, however, in most of the other models 
the mass of the inflaton field at the false vacuum is large compared
with the Hubble mass scale there and in such cases
these discrete modes were found to disappear.

In the case of a single field model, it was argued that there exists yet
another type of discrete supercurvature
modes\cite{HAMA,Garriga,Bellido,YST95}.
 They correspond to the fluctuations of the bubble-wall. 
A peculiar nature of these modes is
that they can be regarded as tensor-type modes which are spatially
transverse-traceless by a suitable coordinate transformation.
It was not clear if these fluctuations contribute also in the case of a
two field model, since the wall fluctuations are associated with the
scalar field which mediates the false vacuum decay but not with the
inflaton field inside the bubble.

Recently, a method to calculate the spectrum of quantized gravitational
waves has been developed\cite{TS97,BucCoh}.
Then it has been shown that the spectrum shows
infrared divergence on pure de Sitter space but it disappears once the
existence of the bubble wall is appropriately taken into account.
It has been also shown that the spectrum of gravitational wave modes 
is continuous and there exists no discrete modes\cite{TS97}.
At the same time, it has been shown that the discrete bubble-wall
fluctuation modes, which had been found in previous analyses,
 cease to exist once the coupling between the scalar
field and the metric 
perturbation is correctly taken into account\cite{TS97}.
Thus the situation has become rather confusing.

In this paper, using the spectrum of gravitational wave modes
obtained in Ref.~\cite{TS97}, we calculate the tensor CMB anisotropy on
large angular scales. The tensor CMB anisotropy in one-bubble inflation
models has been calculated recently by Hu and White\cite{HuWhi}.
However, the gravitational wave spectrum they have assumed for their
calculations seems to be rather ad hoc and they have not discussed
constraints on the model parameters. Here we investigate the dependence
of the tensor CMB anisotropy on the model parameters in detail.
In doing so, we resolve the problem of
the wall fluctuation modes. We then discuss observational constraints
on the parameters of one-bubble inflation models.
 The paper is organized as follows.
In section 2, we review the solution of the bubble configuration
with gravity under the thin-wall approximation and consider the wall
fluctuation modes. We find the wall fluctuation modes are transmuted
into a part of the continuous spectrum of the gravitational wave modes
when the gravitational coupling is taken into account, thus resolving
the problem of the wall fluctuation modes. In section 3, we calculate
the spectrum of the tensor CMB anisotropy and show that it can be
cleanly separated into two parts; one due to the wall fluctuation modes
and the other due to the usual gravitational wave modes.
In section 4, using the results of section 3,
we consider constraints on the model parameters from the
observed CMB anisotropy by COBE. We find the allowable region of the
parameter space is relatively small but still large enough to maintain
the viability of one-bubble inflation models for an open universe.
Finally, section 5 is devoted to summary and discussion.

%%%%%%%%%%%%%%%%%%%%%%%%%%%%%%%%%%%%%%%%%%%%%%%%%%%%%%%%%%%%%%%
\section{Thin-wall bubble and wall fluctuation modes}
In this section, we first review an $O(4)$-symmetric bubble solution
under the thin-wall approximation\cite{Parke,Bellido}
and how the bubble wall fluctuations
are described when the coupling between the scalar-type and tensor-type
perturbations is neglected. Then we
consider vacuum fluctuations of gravitational wave modes and show that 
these vacuum fluctuations exactly coincide with the bubble wall
fluctuations in the limit of small vacuum energy difference between the
true and false vacua or in the limit of small bubble radius.

An $O(4)$-symmetric bubble configuration that mediates the false vacuum
decay is described by the Euclidean metric,
\begin{equation}
ds_E^2=d\tau^2+a_E^2(\tau)\left(d\rho^2+\sin^2\rho\,d\Omega^2\right),
\label{Emetric}
\end{equation}
and by the scalar field $\sigma$ which depends only on $\tau$;
$\sigma=\sigma(\tau)$. The Euclidean action for this configuration is
\begin{eqnarray}
S_E&=&2\pi^2\int d\tau\, a_E^3\left[-{3\over\kappa}
\left\{\left({\dot a_E\over a_E}\right)^2+{1\over a_E^2}\right\}
+{1\over2}\dot\sigma^2+V(\sigma)\right]
\nonumber\\
   &=&2\pi^2\int d\tau
  \left[a_E^3\dot\sigma^2-{6\over\kappa}\dot a_E^2a_E\right],
\label{Eaction}
\end{eqnarray}
where $\kappa=8\pi G=8\pi/M_{pl}^2$, 
$\dot{~}=d/d\tau$ and the second equality follows
 from the constraint equation (or Euclidean Friedmann equation),
\begin{equation}
\left({\dot a_E\over a_E}\right)^2-{1\over a_E^2}
={\kappa\over3}\left({1\over2}\dot\sigma^2-V\right).
\end{equation}
Assuming that the bubble wall is infinitesimally thin, we have
\begin{equation}
\dot a_E^2=
\left\{\begin{array}{ll}
 1-\displaystyle{\kappa\over3}V_Fa_E^2
=:1-H_L^2a_E^2&\quad\hbox{in false vacuum},\\
\\
 1-\displaystyle{\kappa\over3}V_Ta_E^2
=:1-H_R^2a_E^2&\quad\hbox{in true vacuum},\\
\end{array}\right.
\end{equation}
where $V_T$ and $V_F$ are the vacuum energies at the true and false vacua,
 respectively.
The above equation can be easily solved to give
\begin{equation}
a_E=\left\{\begin{array}{ll}
 \displaystyle{1\over H_L}
\cos H_L\tilde\tau\,,\quad& 
-\displaystyle{\pi\over2H_L}\leq\tilde\tau<\tilde\tau_W\,,\\
\\
 \displaystyle{1\over H_R}
\cos H_R\tau\,,\quad& \tau_W<\tau\leq\displaystyle{\pi\over2H_R}\,,\\
\end{array}\right.
\label{Esol}
\end{equation}
where $\tilde\tau=\tau-\tau_W+\tilde\tau_W$ and $\tau_W$ is the value of
 $\tau$ at the wall. 
 Note that the continuity of $a_E$ at the wall implies
\begin{equation}
a_E(\tau_W)={1\over H_L}\cos H_L\tilde\tau_W={1\over H_R}\cos H_R\tau_W\,,
\end{equation}
and hence $\cos H_L\tilde\tau_W>\cos H_R\tau_W$ since $H_L>H_R$.
Note also that $\tau_W\,$, $\tilde\tau_W>0$
because the maximum of the scale factor $a_E$ should be on the false
vacuum side if the bubble configuration describes the false vacuum
decay. We should mention, however, that this condition on
$\tau_W$ and $\tilde\tau_W$ comes from the limit of applicability
of our formalism\cite{TS94}.
Hence models which violate this condition are not excluded a priori.
We do not consider such models here simply because we
do not know the outcome.

Inserting the solution (\ref{Esol}) to Eq.~(\ref{Eaction}),
we then obtain the reduced action for the thin-wall bubble:
\begin{equation}
S_{TW}=2\pi^2\left[S_1 R^3-{2\over\kappa}
\left({(1-H_L^2R^2)^{3/2}-1\over H_L^2}
-{(1-H_R^2R^2)^{3/2}-1\over H_R^2}\right)\right],
\end{equation}
where $R$ and $S_1$ are the radius and the surface tension of the wall,
respectively:
\begin{equation}
R:=a_E(\tau_W),\quad S_1:=\int \dot\sigma^2d\tau\,.
\end{equation}
Then the wall radius is determined by putting $dS_{TW}/dR=0$,
which gives
\begin{equation}
{\kappa\over2}RS_1=(1-H_R^2R^2)^{1/2}-(1-H_L^2R^2)^{1/2}.
\label{Req}
\end{equation}
This can be solved for $R$ by elementary algebra. We find
\begin{eqnarray}
R&=&{\kappa S_1\over\sqrt{(H_L^2-H_R^2+({\kappa\over2}S_1)^2)^2
+H_R^2\kappa^2S_1^2}}
\nonumber\\
&=&{3S_1\over\sqrt{(\Delta V+6\pi GS_1^2)^2+24\pi GV_TS_1^2}}\,,
\label{Radius}
\end{eqnarray}
where $\Delta V=V_F-V_T$. We note that one must have 
\begin{equation}
\Delta V>6\pi GS_1^2\,,
\label{DVcond}
\end{equation}
which comes from the condition $\tau_W$, $\tilde\tau_W>0$. 
For our present purpose, however, it is convenient to
return to Eq.~(\ref{Req}) and rewrite it as
\begin{equation}
{\kappa\over2}RS_1
=\Delta s:=\sin H_R\tau_W-\sin H_L\tilde\tau_W\,.
\label{RS1}
\end{equation}
Note that $(\kappa/2)RS_1<1$ for any bubble configuration which
describes the $O(4)$-symmetric false vacuum decay.
One could consider a configuration in which $\tilde\tau_W<0$,
in which case $(\kappa/2)RS_1$ could be greater than unity.
However, such a configuration would not describe the false vacuum
decay as mentioned before.

In order to clarify the parameter-dependence of the bubble
configuration, it is convenient to introduce the following
non-dimensional parameters:
\begin{equation}
\alpha:={\Delta V\over6\pi GS_1^2}\,,\quad
\beta:={V_T\over6\pi GS_1^2}\,.
\label{alpbeta}
\end{equation}
Note that $\alpha>1$ because of the condition (\ref{DVcond}).
In terms of these, we have
\begin{eqnarray}
&&\cos H_R\tau_W={2\sqrt{\beta}\over\sqrt{(\alpha+1)^2+4\beta}}\,,
\quad
\sin H_R\tau_W={\alpha+1\over\sqrt{(\alpha+1)^2+4\beta}}\,,
\nonumber\\
&&\cos H_L\tilde\tau_W
={2\sqrt{\alpha+\beta}\over\sqrt{(\alpha+1)^2+4\beta}}\,,
\quad
\sin H_L\tilde\tau_W={\alpha-1\over\sqrt{(\alpha+1)^2+4\beta}}\,,
\end{eqnarray}
and 
\begin{equation}
{\kappa\over2}RS_1=\Delta s={2\over\sqrt{(\alpha+1)^2+4\beta}}\,.
\end{equation}
We readily see that $\Delta s\to0$ as
$\alpha\to\infty$ ($H_LR=\cos H_L\tilde\tau_W\to0$) or
$\beta\to\infty$ ($\Delta V/V_T\to0$).
Now let us consider the wall fluctuation modes. Assuming that
the coupling of the scalar field perturbation to the metric perturbation
can be neglected, it has been shown that
quantum fluctuations of the bubble wall induced by the false vacuum decay
give rise to the scalar field perturbation on the time constant 
hypersurface of an open universe inside the bubble, and consequently 
give rise to the curvature perturbation on the comoving hypersurface
on which $\sigma$ is constant\cite{Garriga,Bellido,YST96}. 

The background metric for the open universe inside the bubble is
given by analytic continuation of the Euclidean metric (\ref{Emetric})
as
\begin{equation}
ds^2=-dt^2+a^2(t)(dr^2+\sinh^2r\,d\Omega^2)
=-dt^2+a^2(t)\gamma_{ij}dx^idx^j,
\end{equation}
where 
\begin{equation}
t=i\left(\tau-{\pi\over2H_R}\right),\quad
r=i\rho,\quad a(t)={\sinh H_Rt\over H_R}\,,
\end{equation}
and $\gamma_{ij}$ is the metric on the unit hyperboloid.
On these coordinates, the spatial
curvature perturbation due to the wall fluctuation
modes is expressed as
\begin{equation}
{\cal R}_c=\sum_{lm}{\cal R}_W(t){\cal Y}_{2lm}(r,\Omega)\,,
\end{equation}
where ${\cal Y}_{2lm}$ are the spatial harmonics given by
\begin{equation}
{\cal Y}_{2lm}=\sqrt{\Gamma(l+3)\Gamma(l-1)\over2}
{P^{-l-1/2}_{3/2}(\cosh r)\over\sqrt{\sinh r}}Y_{lm}(\Omega).
\end{equation}
The mean quare value of ${\cal R}_W$ has been found to be\cite{YST96}
\begin{equation}
|{\cal R}_W|^2={H_R^2\over RS_1}={\kappa H_R^2\over2\Delta s}\,.
\end{equation}
Note that this diverges for $RS_1\to0$, i.e., in the limit of small vacuum 
energy difference or small wall radius.

An important property of this curvature perturbation is that it can be
regarded as a tensor-type perturbation which is spatially
transverse-traceless\cite{HAMA,Garriga,YST96}.
Namely it is equivalent to the metric perturbation,
\begin{equation}
H_{ij}=-2{\dot a\over a}H_R^{-1}
{\cal R}_W{\cal Y}_{ij}=:H_W{\cal Y}_{ij}\,,
\end{equation}
where $H_{ij}$ is the perturbation of the spatial metric $\gamma_{ij}$,
${\cal Y}_{ij}$ is defined by
\begin{equation}
{\cal Y}_{ij}:={\cal Y}_{|ij}-\gamma_{ij}{\cal Y}\,,
\label{calYij}
\end{equation}
and we have suppressed the indices $\{2lm\}$ of ${\cal Y}_{2lm}$.
The spatial tensor ${\cal Y}_{ij}$ is transverse-traceless and
corresponds to an even parity tensor harmonic with the eigen value
$p^2=0$. 
Thus the bubble wall fluctuation modes can be regarded as
even parity gravitational wave modes which exist discretely at
$p^2=0$. The mean square value of $H_W$ is given by
\begin{equation}
|H_W|^2\to4|{\cal R}_W|^2={2\kappa H_R^2\over\Delta s}\,,
\label{Hwall}
\end{equation}
for $H_Rt\gg1$.

Let us now turn to the quantum fluctuations of gravitational wave modes.
It has been shown that the spectrum of gravitational wave modes consists
of continuous modes only and there appears no supercurvature
 mode\cite{TS97}.
As for even parity $p^2=0$ modes, for which there appears degeneracy
between the scalar-type and tensor-type perturbations, it has been
shown that the discrete wall fluctuation modes cease to exist
once the coupling
between the scalar field perturbation and the gravitational wave
perturbation is fully taken into account\cite{TS97}. 
Then an immediate question is where the wall fluctuation modes go. 
A peculiar feature of the gravitational wave spectrum is that
it diverges in the limit $p\to 0$ in the pure open de Sitter space but
it becomes finite once the presence of the bubble wall is taken into
account\cite{TS97,BucCoh}. 
Thus one suspects if this feature is related to the problem of
wall fluctuation modes. 
Below we show that the wall fluctuation modes are in fact contained
in the continuous spectrum of even parity gravitational wave modes
in the vicinity of $p=0$.

We confine our attention to the even parity gravitational wave
perturbation and express it as
\begin{equation}
H_{ij}=\int_{-\infty}^\infty dp\,\sum_{lm}
U_{plm}(t)Y^{plm}_{ij}(r,\Omega)\,,
\label{Hij}
\end{equation}
where $Y^{plm}_{ij}$ are the even parity
tensor harmonics which are normalized as
\begin{equation}
\int d\Sigma\,\gamma^{ii'}\gamma^{jj'}Y^{plm}_{ij}
\overline{Y^{p'l'm'}_{i'j'}}
=\delta(p-p')\delta_{l,l'}\delta_{m,m'}\,,
\end{equation}
With this normalization, it can be shown that ${\cal Y}_{ij}$ given
by Eq.~(\ref{calYij}) coincides with $Y^{plm}_{ij}|_{p=0}$\cite{TS97}:
\begin{equation}
{\cal Y}_{ij}=Y^{plm}_{ij}|_{p=0}:=Y^{0}_{ij}\,.
\end{equation}
By quantizing $H_{ij}$ and assuming that the state is in the Euclidean
vacuum, the spectrum of $U_{plm}$ has been found to be\cite{TS97}
\begin{equation}
|U_p|^2:=\sum_{\pm}\langle U_{\pm p\, lm}^2\rangle
 ={4\kappa H^2\coth \pi p\over 2p(1+p^2)}(1-y),
\label{spec}
\end{equation}
where
\begin{equation}
1-y= 1-{(\Delta s)^2 \cos bp
     +2p \Delta s\sin bp
      \over(4p^2+(\Delta s)^2)\cosh\pi p}\,,
\end{equation}
with $\Delta s$ being defined in Eq.~(\ref{RS1}) and
\begin{equation}
b:=\ln\left(\displaystyle
{1+\sin H_R\tau_W\over 1-\sin H_R\tau_W}\right)
=\ln{\left(\sqrt{(\alpha+1)^2+4\beta}+\alpha+1\right)^2\over
4\beta}\,.
\end{equation}

In the limit $\Delta s\ll 1$, i.e., $\alpha$ or $\beta$ is large,
assuming $b\alt \Delta s^{-1}$, 
which holds except for an unrealistic case of exponentially small
$\beta$,
the spectrum (\ref{spec}) is sharply peaked around $p=0$ and
can be approximated as
\begin{equation}
 |U_p|^2\approx{2\kappa H_R^2\over\pi}{1\over p^2+(\Delta s/2)^2}\,.
\label{Uapprox}
\end{equation}
Then we can regard the gravitational wave perturbation as 
being dominantly given by the $p=0$ modes only with the mean square
amplitude given by
\begin{equation}
\int_0^\infty dp\,|U_p|^2\approx{2\kappa H_R^2\over\Delta s}\,.
\label{Upint}
\end{equation}
We find this mean square amplitude exactly coincides with that of the
wall fluctuation modes, Eq.~(\ref{Hwall}).
Therefore, since ${\cal Y}_{ij}=Y^{0}_{ij}$, we conclude
that the gravitational wave perturbation in this case is 
actually the one due to the wall fluctuations.
This is a reasonable result. Since the limit of small 
$\Delta s$($=(\kappa/2)RS_1$) can be
regarded as the limit of weak gravity (i.e., $\kappa\to0$), we should
obtain the same result as the one obtained by neglecting the 
gravitational degrees of freedom.\footnote{The form 
of Eq.~(\ref{Uapprox}) suggests that the pole at $p=i\Delta s/2$
may have some physical significance for description of
the wall fluctuation modes.
In fact, a detailed analysis of the behavior of mode functions
in the region outside the lightcone emanating from the origin $t=r=0$
(the region $C$ in Ref.~\cite{YST96} or \cite{TS97}) shows that 
the spacetime with the bubble wall
admits a quasi-normal mode of the metric perturbation at
$p=i\Delta s/2$ which satisfies the purely outgoing boundary condition
there. It will be extremely interesting if the contribution
of the wall fluctuation modes to the spectrum can be described as
a quantum excitation of the quasi-normal mode.}

\begin{figure}
\epsfysize=8cm
\vspace{-2.5cm}
\centerline{\epsfbox{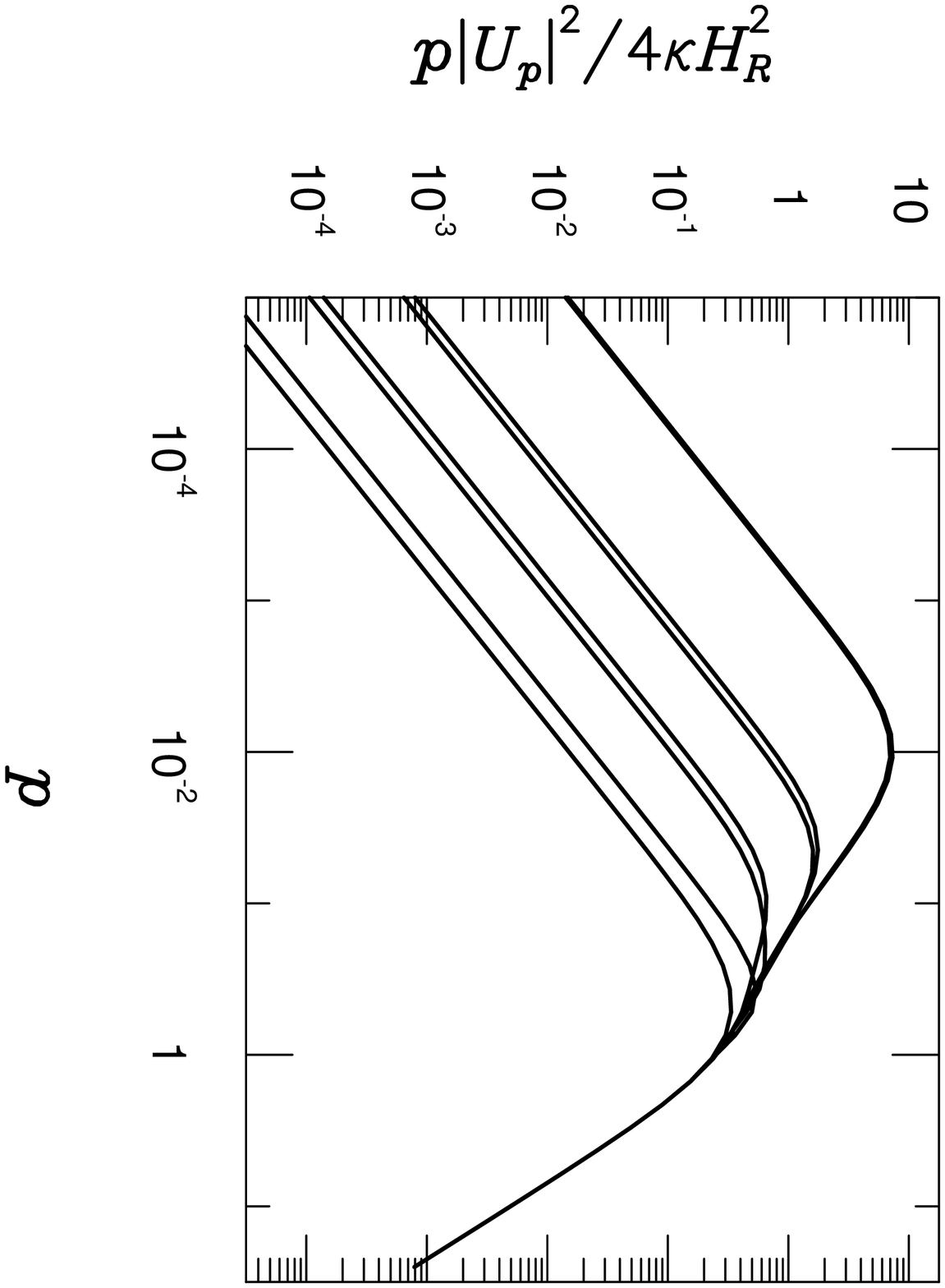}\hspace{2cm}}
\end{figure}
\vspace{-5mm}
\noindent{\small
FIG.~1.
The gravitational wave spectra for various model parameters.
The lines show, from top to bottom,
the cases $(\alpha,\beta)=(10^2,10^2)$, $(10^2,1)$,
$(10^2,10^{-2})$, $(10,10^2)$, $(1,10^2)$, $(10,1)$, $(10,10^{-2})$,
$(1,1)$ and $(1,10^{-2})$. The first three cases
degenerate into a single line in the figure.}
\vspace{5mm}

Here we comment on an important fact. In the previous analyses of the
wall fluctuation modes\cite{Garriga,Bellido,YST96},
in which the coupling to gravity was neglected,
it was not clear if these modes would contribute to the temperature
anisotropy in the case of two field models of open inflation.
The reason is that the wall fluctuations are associated with a scalar
field which causes the false vacuum decay but not with another
scalar field which drives the subsequent inflation, and the
fluctuations of the latter field are thought to be responsible
for the curvature perturbation of the universe.
Now it has become clear that the wall fluctuations {\it do\/}
 contribute to the temperature anisotropy even in the case of
two field models. This implies that the wall fluctuation modes
give rise to a constraint not only on a single field model but also on
a two field model.

We also note that $|U_p|^2p^3\to2\kappa H_R^2$ for $p\gg1$.
Hence the usual scale-invariant spectrum is recovered on scales
smaller than the curvature scale. Therefore the spectrum consists of
two distinguishable parts. What happens in the present case is
that the part due to the wall fluctuation modes dominate over 
the scale-invariant part by a factor $1/\Delta s\gg 1$. 
Then it is natural to speculate that the gravitational wave spectrum
can be decomposed into two parts even if $\Delta s$ is not small; 
one due to the wall fluctuations and the other due to the usual
vacuum fluctuations. As a supporting evidence, we show the spectrum
$|U_p|^2$ for various choices of the model parameters in Fig.~1.
One sees that all the spectra are identical for $p\agt1$ while they
exhibits similar behavior except for the position and the amplitude 
of the maximum for $p\alt1$.
 In the next section, we will
evaluate the large angle CMB anisotropy due to the tensor modes and
 show that this speculation indeed turns out to be true.

%%%%%%%%%%%%%%%%%%%%%%%%%%%%%%%%%%%%%%%%%%%%%%%%%%%%%%%%%%%%%%%
\section{large angle tensor CMB anisotropy}

The CMB temperature anisotropy due to a tensor-type perturbation is
given by
\begin{eqnarray}
{\delta T\over T}(\gamma^i)
&=&-{1\over2}\int_{\eta_{LSS}}^{\eta_0}
d\eta\, H'_{ij}(\eta,x^i(\eta))\gamma^i\gamma^j
\nonumber\\
&=&-{1\over2}\int_{\eta_{LSS}}^{\eta_0}
d\eta\, H'_{rr}(\eta,x^i(\eta))\,,
\end{eqnarray}
where $\eta$ is the conformal time; $d\eta=dt/a(t)$, 
$H_{ij}$ is the transverse-traceless metric perturbation
as given by Eq.~(\ref{Hij}), $H_{rr}$ is the 
$(r,r)$-component of $H_{ij}$,
$H'_{ij}$ is the $\eta$ derivative of $H_{ij}(\eta,x^i)$,
$x^i=x^i(\eta)$ is the photon trajectory, $\gamma^i$ is the unit
vector along the observer's line of sight and we have taken the origin
$r=0$ as the position of the observer.
Since $H_{rr}$ vanishes for the odd parity modes, only the even
parity modes contribute to the temperature anisotropy.
The conformal time of the last scattering surface, $\eta_{LSS}$, and
that at present, $\eta_0$, are given respectively as
\begin{equation}
\eta_{LSS}=2\,{\rm arccosh}\sqrt{1+{\Omega_0^{-1}-1\over1+z_{LSS}}}\,,
\quad
\eta_{0}=2\,{\rm arccosh}\sqrt{\Omega_0^{-1}}\,,
\end{equation}
where $z_{LSS}$ is the redshift of the last scattering surface.
For simplicity, we adopt the value $z_{LSS}=1100$ in the following
calculations, but the results are insensitive to different choices
of $z_{LSS}$.

The evolution equation for $U_{plm}$ is given in terms of the
conformal time as\cite{KS}
\begin{equation}
U_{plm}''+2{a'\over a}U_{plm}'+(p^2+1)U_{plm}=0\,.
\end{equation}
Since we are interested in the anisotropy on large angular scales,
we consider only those modes that come inside the Hubble horizon
after the universe becomes matter-dominated. The scale factor in the 
matter-dominated universe is given by $a(\eta)=\cosh\eta-1$.
Then with the initial condition that $U_{plm}$ approaches a constant as
$\eta\to0$, the above equation can be solved exactly to give
\begin{equation}
U_{plm}(\eta)
=U_{plm}(0){3\left(\displaystyle\cosh{\eta\over2}{\sin p\eta\over2p}
-\sinh{\eta\over2}\cos p\eta\right)
\over\displaystyle(1+4p^2)\sinh^3{\eta\over2}}
=:U_{plm}(0)G_p(\eta),
\label{Usol}
\end{equation}
where $U_{plm}(0)$ has the spectrum given by Eq.~(\ref{spec}).
Then the temperature anisotropy is expressed as
\begin{equation}
\left({\delta T\over T}\right)_{p,l}Y_{lm}
=-{1\over2}U_{plm}(0)\int_{\eta_{LSS}}^{\eta_0}d\eta\,
G_p'(\eta)Y^{plm}_{rr}(\eta_0-\eta,\Omega)\,,
\end{equation}
where we have decomposed the anisotropy as
\begin{equation}
\left({\delta T\over T}\right)=
\int_0^\infty dp\,\sum_{lm}\left({\delta T\over T}\right)_{p,l}Y_{lm}\,.
\end{equation}
The explicit form of $Y^{plm}_{rr}$ is
\begin{equation}
Y^{plm}_{rr}=\sqrt{(l-1)l(l+1)(l+2)\over2(1+p^2)}\,
\left|{\Gamma(l+1+ip)\over\Gamma(1+ip)}\right|\,
{P^{-l-1/2}_{ip-1/2}(\cosh r)\over\sinh^{5/2}r}\,Y_{lm}(\Omega),
\end{equation}
where $P^\mu_\nu$ is the Legendre function of the first kind.

As customarily done, we describe the temperature anisotropy in terms of
the multipole moments $C_l$ of the temperature autocorrelation function:
\begin{equation}
C(\theta):=\left\langle{\delta T\over T}(\gamma^i)
{\delta T\over T}(\tilde\gamma^i)\right\rangle
={1\over4\pi}\sum_l(2l+1)C_lP_l(\cos\theta)\,,
\end{equation}
where $\cos\theta=\gamma_{ij}\gamma^i\tilde\gamma^j$. Then the moment for
the tensor-type perturbation, $C_l^{(T)}$, is given by
\begin{eqnarray}
C_l^{(T)}&=&\int_0^\infty dp
\left\langle\left({\delta T\over T}\right)_{p,l}^2\right\rangle
\nonumber\\
&=&{(l-1)l(l+1)(l+2)\over8}\int_0^\infty dp\,{|U_p|^2\over1+p^2}\,
\left|{\Gamma(l+1+ip)\over\Gamma(1+ip)}\right|^2
\left|\int_{\eta_{LSS}}^{\eta_0}d\eta\, G_p'(\eta)\,
{P^{-l-1/2}_{ip-1/2}(\cosh r)\over\sinh^{5/2}r}\right|^2\,,
\label{Cell}
\end{eqnarray}
where $r=\eta_0-\eta$.

\begin{figure}
\epsfysize=8cm
\vspace*{-2.5cm}
\centerline{\epsfbox{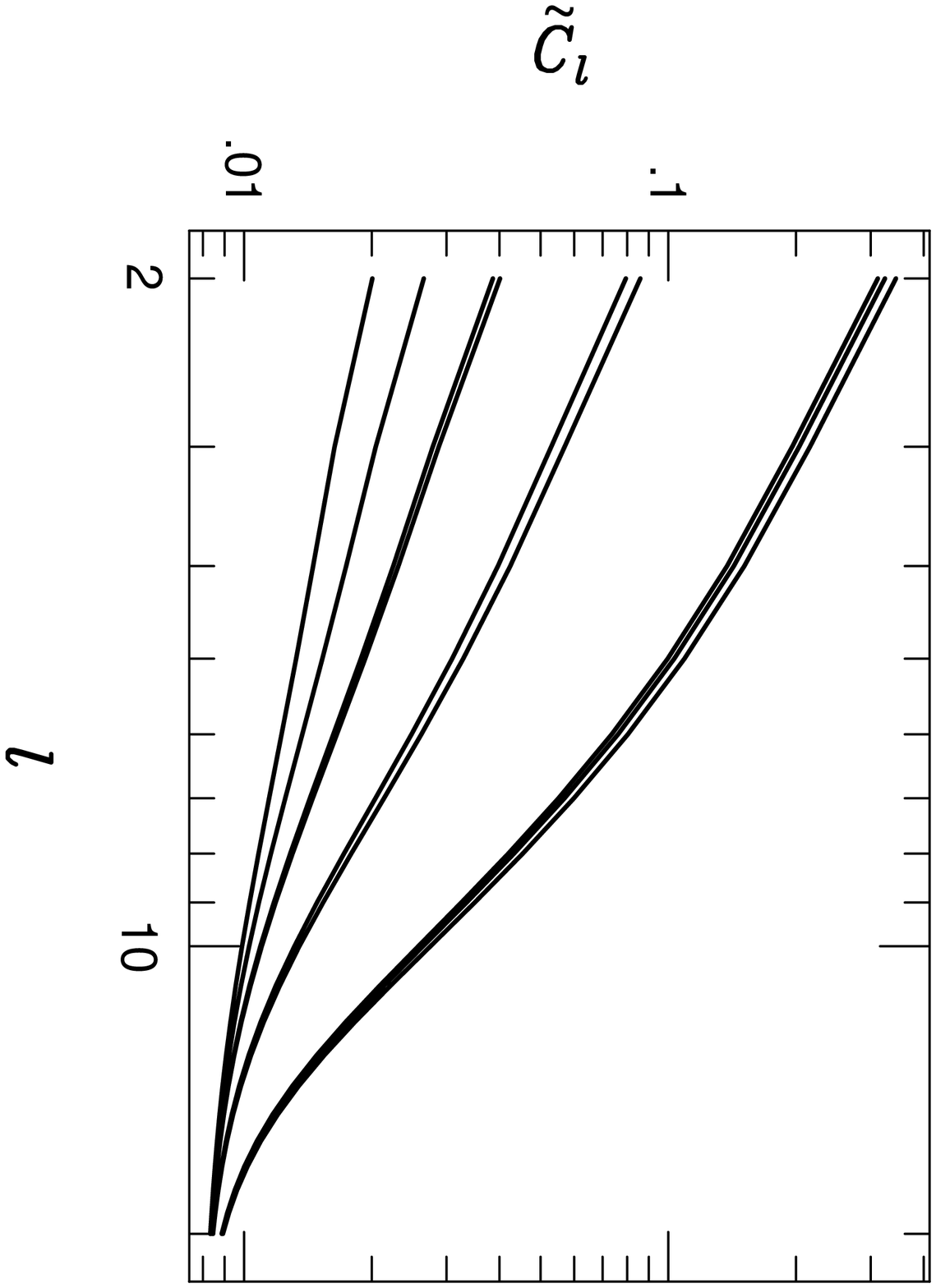}\hspace{3.0cm}\epsfysize=8cm
   \epsfbox{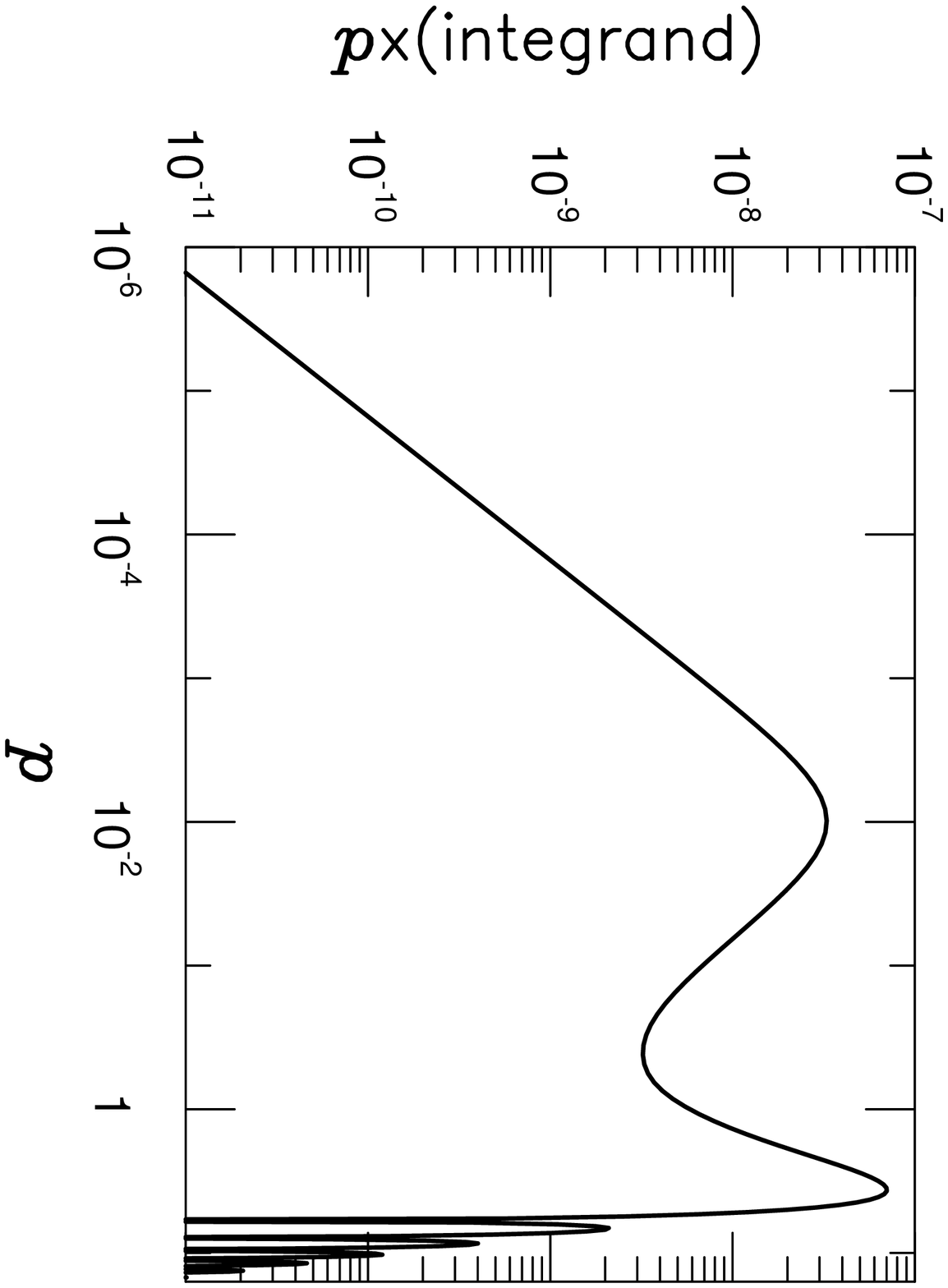}\hspace{2.5cm}}
\end{figure}
\vspace{-5mm}
\centerline{
\begin{minipage}[t]{8.5cm}{
\small 
FIG.~2. 
The normalized tensor CMB anisotropy spectra for various model
parameters in the universe with $\Omega_0=0.3$.
 See the text for the normalization.
The lines show, from top to bottom,
the cases $(\alpha,\beta)=(10^2,10^2)$, $(10^2,1)$,
$(10^2,10^{-2})$, $(10,10^2)$, $(1,10^2)$, $(10,1)$, $(10,10^{-2})$,
$(1,1)$ and $(1,10^{-2})$, where the first three cases, the fourth and
fifth cases, and the sixth and seventh cases, respectively,
are almost degenerate.}
\end{minipage}
\hfill
\begin{minipage}[t]{8.5cm}{
\small
FIG.~3. 
The contribution to
the CMB anisotropy integral from each logarithmic interval of $p$ 
in the case of $(\alpha,\beta)=(10^2,1)$ for $l=10$ and $\Omega_0=0.3$.
The wide peak around $p\sim10^{-2}$ is due to the wall fluctuation
modes.}
\end{minipage}}
\vspace{5mm}

The anisotropy spectra $\tilde C_l:=l(l+1)C_l^{(T)}/(4\kappa H_R^2)$
for various different model parameters in the
$\Omega_0=0.3$ universe are shown in Fig.~2. We see that all the spectra
converge down to the same constant value as $l$ increases, which presumably
corresponds to the part due to the conventional scale-invariant
spectrum. This suggests that the enhancement on smaller $l$ indeed
represents the contribution from the wall fluctuation modes. 
A further evidence that the spectrum can be decomposed into the two
parts is found by inspecting the behavior of the integrand of
Eq.~(\ref{Cell}) as a function of $p$. We show in Fig.~3 the integrand
times $p$ for $l=10$ in the case of the model parameters $\alpha=10^2$
and $\beta=1$.
The wide peak on the side of smaller $p$ comes from the peak
in the original spectrum of gravitational wave modes $|U_p|^2$ which is
due to the wall fluctuation modes, while the peak on the side of larger
$p$ arises from the coherence between the oscillations of 
$P^{-l-1/2}_{ip-1/2}(\cosh r)$ and $G'_p(\eta)$, 
which may be regarded as the
contribution from the scale-invariant part of the gravitational wave
spectrum.
To examine this conjecture, we pose the following ansatz as the form
of the CMB spectrum:
\begin{equation}
l(l+1)C_l^{(T)}=4\kappa H_R^2\left(A\tilde C_l^{(W)}
+\tilde C_l^{(R)}\right)\,,
\label{Clform}
\end{equation}
where, except for the dependence on $\Omega_0$,
 $\tilde C_l^{(W)}$ and $\tilde C_l^{(R)}$ are assumed to be
model-independent
and represent contributions from the wall fluctuation modes and the 
residual continuous spectrum of gravitational wave modes, respectively.
Thus the only model-dependent part is the factor $A$, which we 
assume to be given by the total
integral of the gravitational wave spectrum:
\begin{equation}
A={1\over4\kappa H_R^2}\int_0^\infty dp|U_p|^2\,.
\end{equation}
Note that $A\approx 1/(2\Delta s)$ for $\Delta s\ll 1$
from Eq.~(\ref{Upint}). 
It should be also kept in mind that 
there always exists freedom in the choice of $A$ by a constant shift:
Replacing $A$ with $A+C$ where $C$ is a model-independent constant
will be equally valid if Eq.~(\ref{Clform}) holds. 
The only reason for our choice of $A$ is that it is simplest and seems
most natural.

 If the above ansatz (\ref{Clform}) works,
$\tilde C_l^{(W)}$ will be given by the spectrum for the case $\Delta
s\to0$, which we show in Fig.~4 for various values of $\Omega_0$. 
As expected, we find $\tilde C_l^{(W)}$ precisely agrees with the
spectrum due to the wall fluctuation modes calculated previously
(See Fig.~4 in \cite{YST96}).
Then the residual part $C_l^{(R)}$ can be easily calculated. We plot the
results for various cases of model parameters in Fig.~5. We find the
parameter-dependence is surprisingly small, which is exhibited in the
narrow thickness of each curve for each value of $\Omega_0$. Thus we
conclude that the tensor CMB anisotropy can be
decomposed into the part due to wall fluctuation modes and
the part due to continuous gravitational wave modes and the
model-dependence is conveniently characterized by the factor $A$ which
is the total integral of the gravitational wave spectrum.

\begin{figure}
\vspace*{-2.5cm}
\centerline{\epsfysize=8cm
\epsfbox{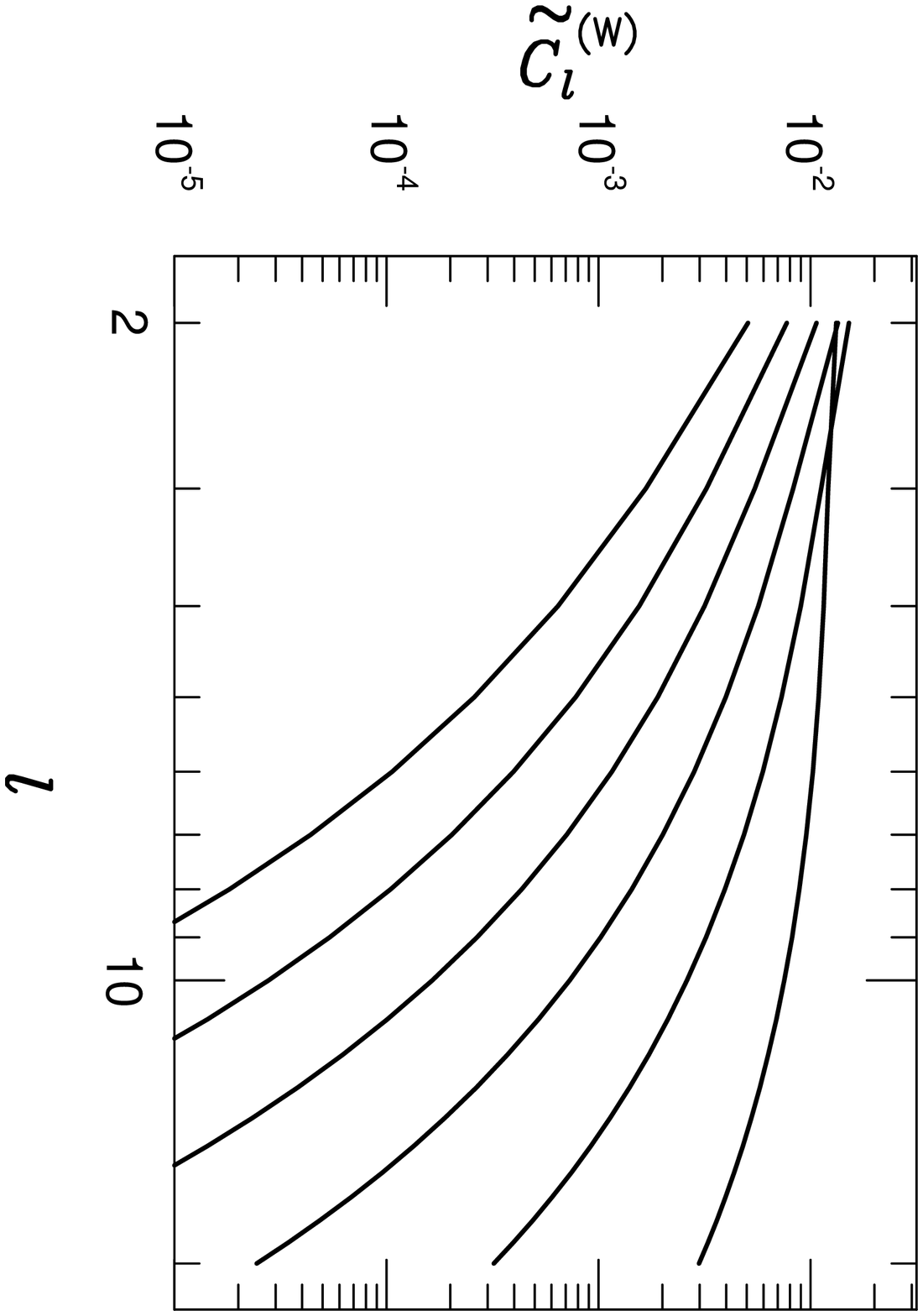}\hspace{3.0cm}
\epsfysize=8cm
\epsfbox{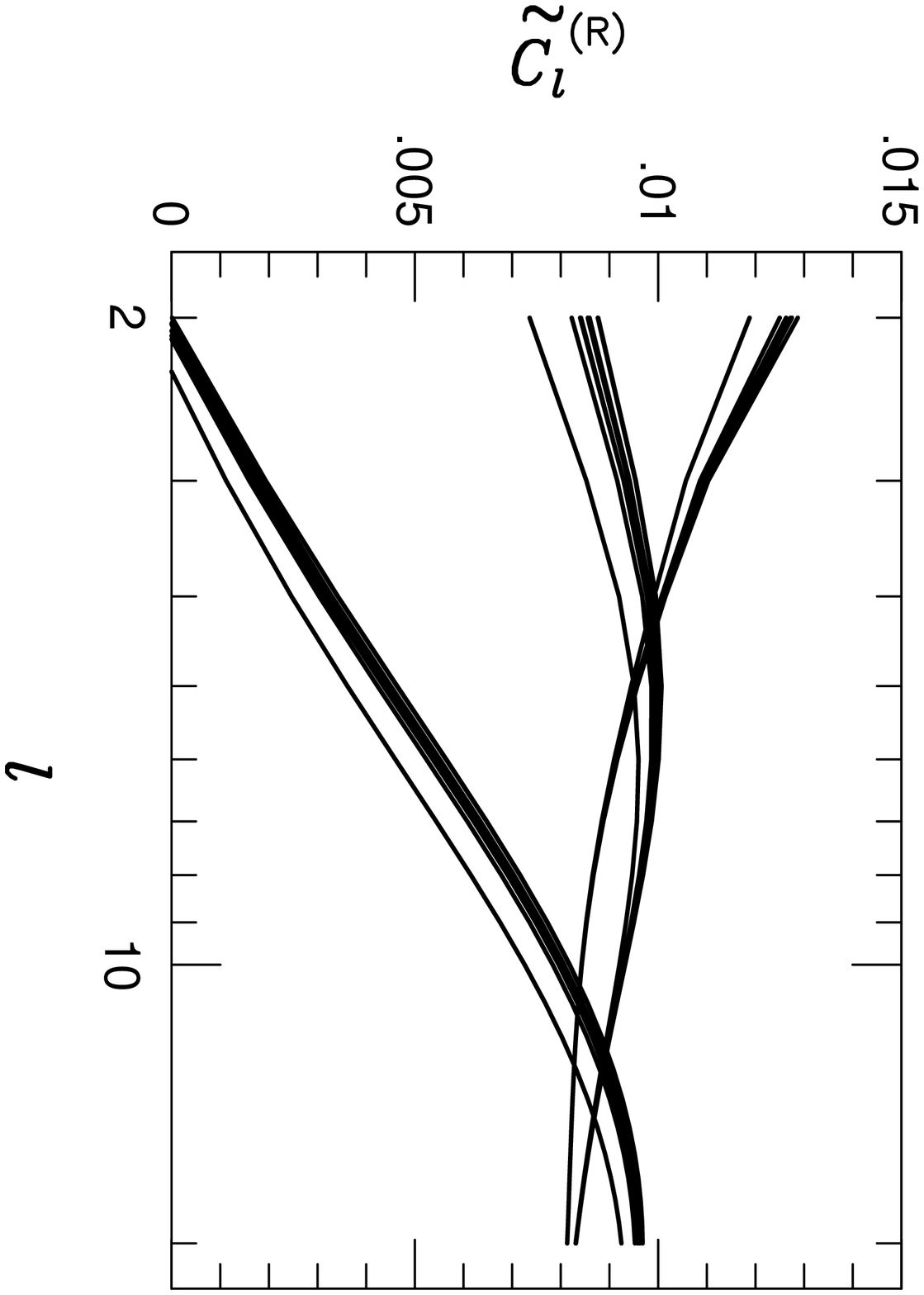} 
\hspace{2.5cm}}
\end{figure}
\vspace{-5mm}
\centerline{
\begin{minipage}[t]{8.5cm}{
\small 
FIG.~4. 
The normalized tensor CMB anisotropy in the limit
$\Delta s\to0$. The lines show, from top to bottom (at larger $l$),
the cases of the universe with $\Omega_0=0.1$, 0.2, 0.3, 0.4, 0.5 and
0.6.}
\end{minipage}
\hfill
\begin{minipage}[t]{8.5cm}{
\small 
FIG.~5. 
The residual, almost model-independent part of the tensor CMB
anisotropy spectra. The top bunch of lines at $l=2$ are the universe
with $\Omega_0=0.5$, the second bunch with $\Omega_0=0.3$ and the third
bunch with $\Omega_0=0.1$. Each bunch contains the nine cases of
$\alpha=10^{-2}$, 1, $10^2$ and $\beta=10^{-2}$, 1, $10^2$,
respectively. The bottom line in each bunch that is slightly away from
the other ones corresponds to the case $\alpha=\beta=1$.
}
\end{minipage}}

\section{Constraints on the model parameters}

In this section, we consider constraints on the model parameters by
comparing the tensor CMB anisotropy calculated in the previous section
with the scalar CMB anisotropy which is due to the quantum fluctuations
of the inflaton field. 

For definiteness, we consider either a single
field or two field model of open inflation in which the potential of the
inflaton field has the form $V=\lambda \phi^{2n}$ inside the bubble.
We also assume that the mass of the inflaton field at the false vacuum
is much greater than the Hubble parameter there so that there is no
contribution from the de Sitter supercurvature modes to the scalar CMB
anisotropy. 
Then it has been shown that the CMB anisotropy spectrum is
almost model-independent and indistinguishable from the one
for the Bunch-Davies vacuum or the conformal vacuum\cite{YST96}. 
Therefore, as a
reference spectrum for the scalar CMB anisotropy, we take the one
calculated for the case of the conformal vacuum. We express it as
\begin{equation}
l(l+1)C_l^{(S)}=\left({3H_R^2\over5\dot\phi}\right)^2\tilde C_l^{(S)}\,.
\label{Clscalar}
\end{equation}
The normalized spectra $\tilde C_l^{(S)}$ for various values of
$\Omega_0$ are plotted in Fig.~6.

As seen from Fig.~4, the anisotropy spectrum due to the wall
fluctuation modes rises rather steeply towards lower $l$. Hence if it
dominates over the scalar CMB anisotropy at $l\alt10$, it will
contradicts with the anisotropy spectrum observed by COBE which 
is relatively flat there\cite{COBE}. Hence we consider the ratio,
\begin{equation}
r_l:={l(l+1)C_l^{(W)}\over l(l+1)C_l^{(S)}|_{l=10}}
=\left({3H_R^2\over5\dot\phi}\right)^{-2}
4\kappa H_R^2A{\tilde C_l^{(W)}\over\tilde C_{10}^{(S)}}
={25\over18\pi}{A\over\zeta^2}
{\tilde C_l^{(W)}\over\tilde C_{10}^{(S)}}\,,
\label{WSratio}
\end{equation}
where 
\begin{equation}
\zeta:={V\over V'M_{pl}}={M_{pl}\over2n\phi}
\approx\sqrt{N(\phi)\over8n\pi}\,,
\end{equation}
and $N(\phi)$ is the e-folding number of expansion from the time the
inflaton has the value $\phi$ until the end of inflation.
In the present case, a relevant value of $N$ will be $\sim60$ as we are
interested in the present Hubble horizon scale. Then $\zeta^2\sim3$ for
$n=1$. The ratio $\tilde C_l^{(W)}/\tilde C_{10}^{(S)}$ is a factor
which can be calculated model-independently. The result is shown in
Fig.~7 for $\Omega_0=0.1\sim0.6$. 
Then a bound on the value
of $r_l$ translates to that on $A$, which in turn constrains the model
parameters. Although it is hard to tell precisely the maximum
allowable value of $r_l$, we tentatively require $r_l<1$ as a 
conservative bound. Then we have
\begin{equation}
A\alt 7\left({\zeta^2\over3}\right)
\left({\tilde C_l^{(W)}\over\tilde C_{10}^{(S)}}\right)^{-1}\,.
\label{Abound}
\end{equation}

\begin{figure}
\vspace*{-2.5cm}
\centerline{\epsfysize=8cm\epsfbox{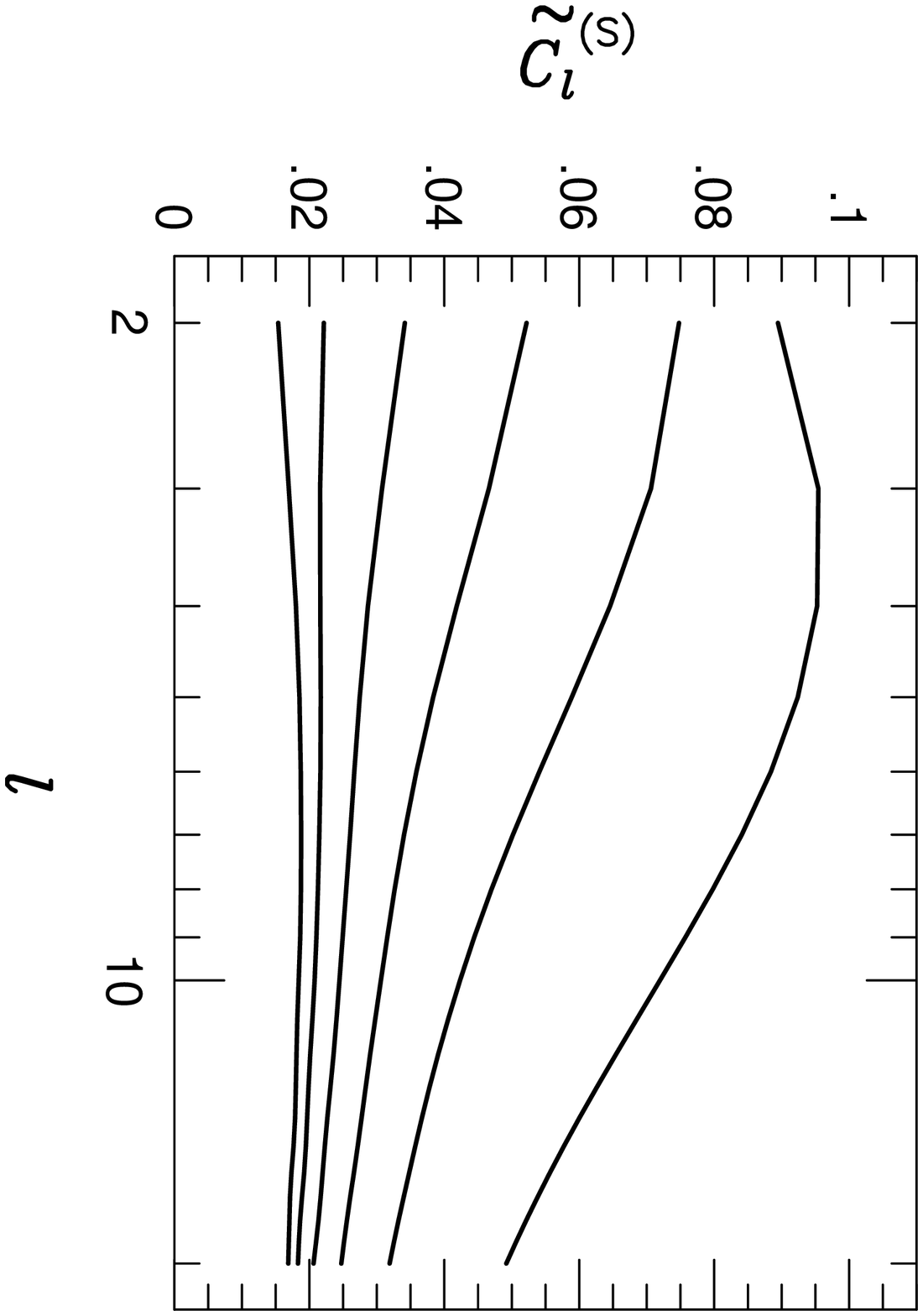}\hspace{3.0cm}
\epsfysize=8cm\epsfbox{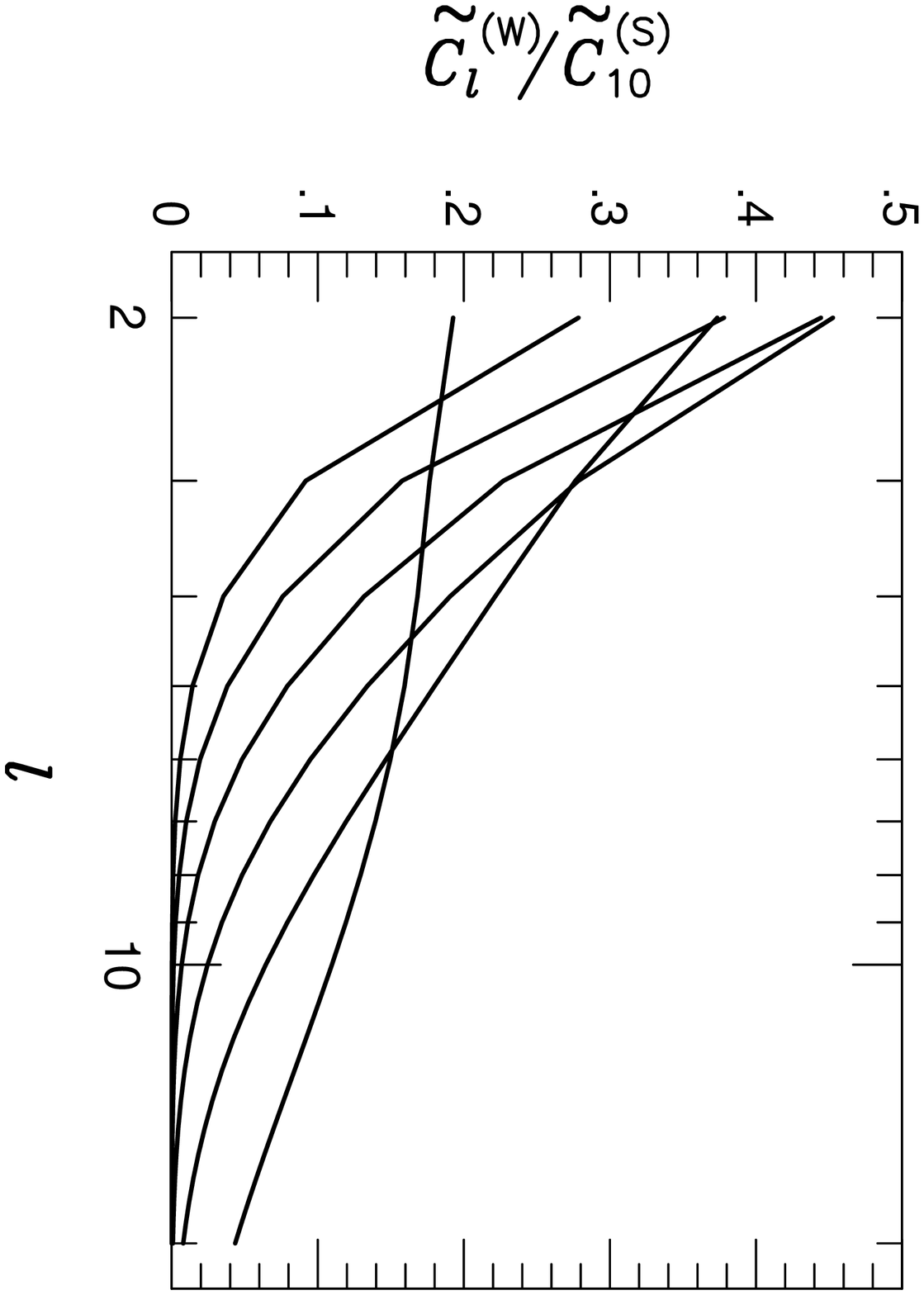}\hspace{2.5cm}}
\end{figure}
\vspace{-5mm}
\centerline{
\begin{minipage}[t]{8.5cm}{
\small 
FIG.~6. 
The normalized scalar CMB anisotropy spectra for
various values of $\Omega_0$. The lines show, from top to bottom,
the cases of $\Omega_0=0.1$, 0.2, 0.3, 0.4, 0.5 and 0.6.
See Eq.~(\ref{Clscalar}) for the normalization.}
\end{minipage}
\hfill
\begin{minipage}[t]{8.5cm}{
\small 
FIG.~7. 
The ratio of the normalized CMB 
spectrum due to wall fluctuations to the $l=10$ component of the
normalized scalar CMB anisotropy for various $\Omega_0$.
The lines show, from the top right to the bottom left,
the cases of $\Omega_0=0.1$, 0.2, 0.3, 0.4, 0.5 and 0.6.}
\end{minipage}}
\vspace{5mm}

A contour plot of $A$ on the $(\alpha,\beta)$-plane is shown in Fig.~8,
where $\alpha$ and $\beta$ are the non-dimensional parameters of a model
defined in Eqs.~(\ref{alpbeta}).  Then taking account of the ratio
$\tilde C_l^{(W)}/\tilde C_{10}^{(S)}$ shown in Fig.~7, we find
the model parameters are most severely constrained when
$\Omega_0=0.3\sim0.4$. In this case, the constraint (\ref{Abound}) gives
$\alpha\alt70$ and $\beta\alt10^3$. On the other hand, the constraint is
weakest when $\Omega_0=0.1$, for which we obtain $\alpha\alt140$ and
$\beta\alt3\times10^3$. In any case, we conclude that
the constraint is not too tight but
it is necessary to tune the model parameters to some degree.

\begin{figure}
\vspace*{-2.5cm}
\centerline{\epsfysize=9cm
\epsfbox{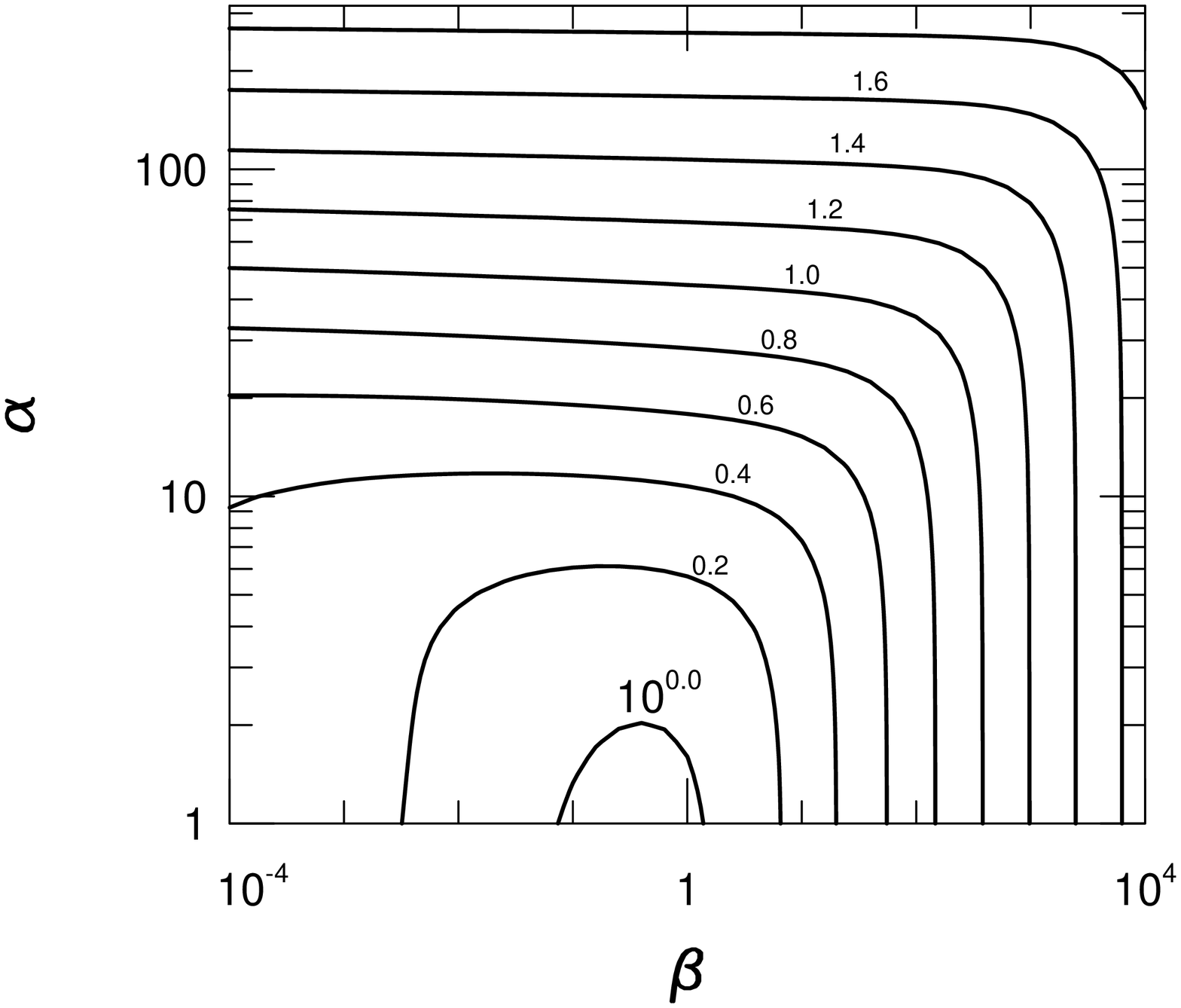}\hspace{2.0cm}
\epsfysize=9cm
\epsfbox{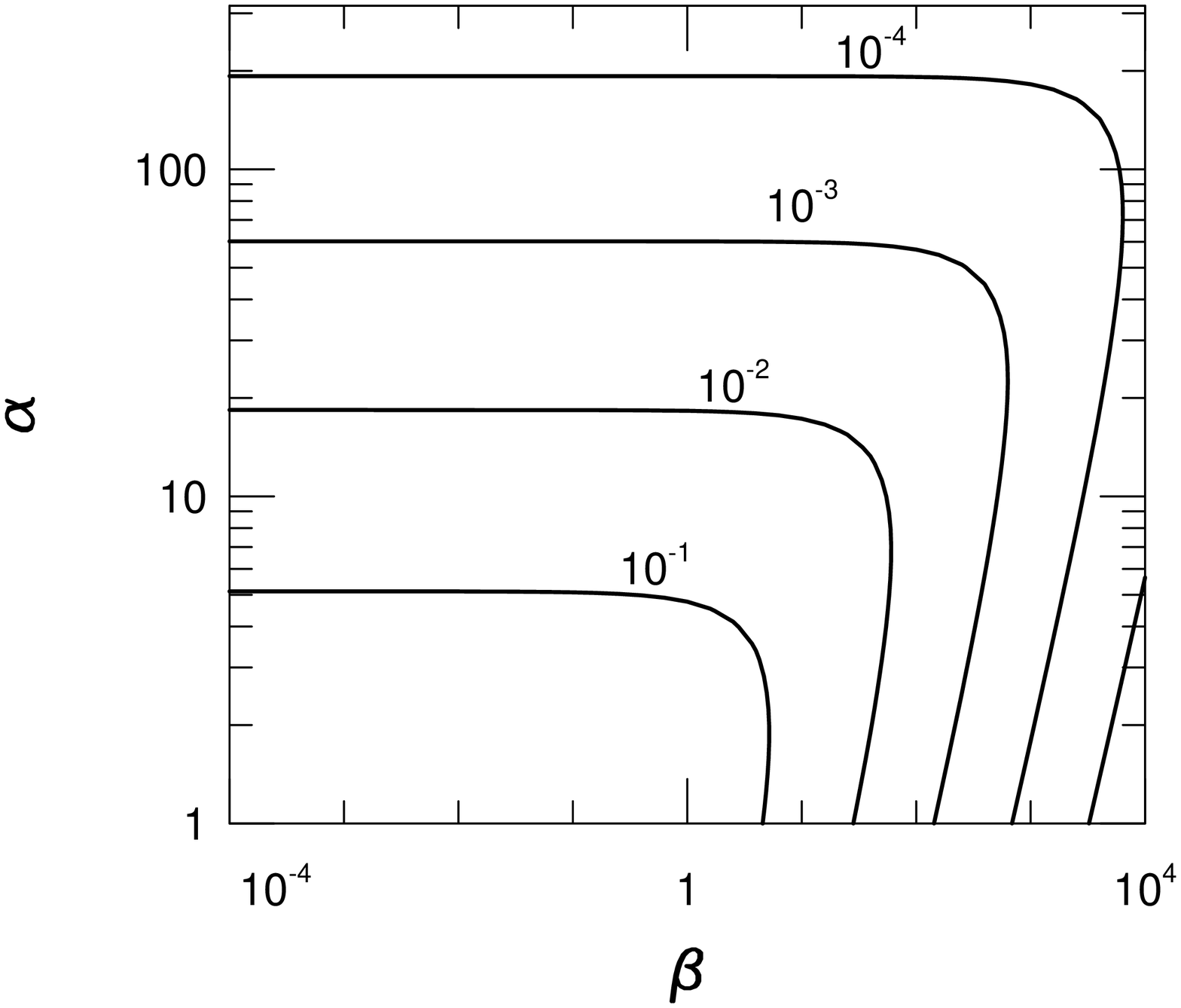}\hspace{2.cm}}
\end{figure}
\vspace{-5mm}
\centerline{
\begin{minipage}[t]{8.5cm}{
\small 
FIG.~8. 
A contour plot of the model-dependent factor $A$
on the $(\alpha,\beta)$ plane. The contours are drawn in every
factor of $10^{0.2}\approx1.58$.} 
\end{minipage}
\hfill
\begin{minipage}[t]{8.5cm}{
\small 
FIG.~9. 
A contour plot of the function $f(\alpha,\beta)$.} 
\end{minipage}}
\vspace{5mm}

Finally, we consider another important condition for a one-bubble open
inflation model to be successful. Since the nucleated bubbles should
not collide with each other, the nucleation rate must be exponentially
suppressed. In other words, the exponent $B$ of the bubble nucleation
rate should be much greater than unity. In the present case, $B$ is
given by the action for the $O(4)$-symmetric bubble minus the action for
the purely false vacuum configuration. In terms of $\alpha$ and $\beta$,
it is expressed as
\begin{equation}
B={M_{pl}^4\over\Delta V}f(\alpha,\beta)\,,
\end{equation}
where
\begin{equation}
f(\alpha,\beta)={3\alpha\over16\gamma^3}
\left[8-{(\alpha-1)^3\over\alpha+\beta}+{(\alpha+1)^3\over\beta}
+{\gamma^3\over\alpha+\beta}-{\gamma^3\over\beta}\right]\,,
\end{equation}
and $\gamma=\sqrt{(\alpha+1)^2+4\beta}$. A contour plot of $f$ is given
in Fig.~9. We find that $f\agt10^{-5}$ in the whole region of our
interest. Thus we have $B\gg1$ as long as $\Delta V\ll10^{-5}M_{pl}^4$,
which is easily satisfied for any reasonable choice of $\Delta V$.

\section{Summary and discussion}

We have calculated the tensor CMB anisotropy on large angular scales in
the one-bubble open inflationary universe scenario. We have found that
the wall fluctuation modes, which were previously regarded as
discrete supercurvature modes of the tunneling scalar field, are
transmuted into a part of continuous spectrum of the gravitational wave
modes when the coupling between the scalar field and the metric
perturbations is correctly taken into account and the resulting CMB
anisotropy is identical in the limit of weak gravity. We have then
shown that the tensor CMB anisotropy spectrum
can be conveniently decomposed into two parts; one due to the wall
fluctuation modes and the other due to the usual gravitational wave
modes, even in the case when the gravitational coupling is
non-negligible.

We have then considered constraints on the model
parameters by comparing the calculated tensor CMB
anisotropy with the scalar CMB anisotropy. Assuming a typical chaotic
inflation type potential for the inflaton field inside the bubble,
and requiring that the tensor
anisotropy does not dominate the spectrum at $l\alt10$, we have found
that the allowable region in the parameter space is large enough,
though some tuning of the parameters is necessary. Specifically, we
have found that the constraints are most stringent for the universe with
$\Omega_0=0.3\sim0.4$, for which we have obtained
$\alpha=\Delta V/(6\pi GS_1^2)\alt70$ and 
$\beta=V_T/(6\pi GS_1^2)\alt10^3$. 
The condition on the parameter $\beta$
is not tight at all. However,
the parameter $\alpha$ is relatively tightly bound because we must
have $\alpha>1$, which comes from the existence condition of the 
bubble configuration that describes false vacuum decay.

As noted before, however, it is worthwhile to keep in mind that 
the violation of the condition $\alpha>1$ does not directly imply
a failure of the model. We have excluded such models simply
because we cannot predict the outcome within the present formalism.
In this connection, we mention an interesting possibility.
Namely, if one considers a scenario in which the
universe is created with a bubble from nothing, just like the creation
of a universe discussed in the quantum cosmological
context\cite{HarHaw,Vil}, the homogeneity and isotropy of the
classical background universe inside the bubble will be guaranteed
irrespective of the value of $\alpha$. In this case, 
$\alpha$ can take any positive value as long as $\alpha\alt70$.

It has been pointed out that the scalar and tensor contributions to the
CMB anisotropy can be separated by measuring the polarization pattern of
CMB\cite{KaKoSt,SelZal}. 
These works assume the plane wave decomposition of
the metric perturbation. However, it will not be valid for the present
case in which the tensor contribution is dominated by the wall
fluctuation modes whose coherence scale is of 
the order of curvature scale. It is a challenging issue to formulate
a method to describe the CMB polarization due to the curvature scale
modes and to see if there appears a feature characteristic to the
one-bubble inflationary scenario.

Finally, we should mention that we have assumed the inflaton potential
to be of chaotic inflation type inside the bubble, which is of course
not the only possibility. It is interesting to see how the other
types of potential models are constrained by the tensor CMB anisotropy.

\begin{center}
{\bf Acknowledgments}
\end{center}
Y.Y. thanks Prof. S. Ikeuchi for his continuous encouragement. 
This work  was supported in part by Monbusho Grant-in-Aid for
Scientific Research No.~07304033.

\end{document}